# Evidence for gapless quantum spin liquid in a honeycomb lattice

Chengpeng Tu[1], Dongzhe Dai[1], Xu Zhang[1], Chengcheng Zhao[1], Xiaobo Jin[1], Bin Gao[2], Tong Chen[2], Pengcheng Dai[2], and Shiyan Li[1,3,4*]

[1]*State Key Laboratory of Surface Physics, and Department of Physics, Fudan University, Shanghai 200438, China*

[2]*Department of Physics and Astronomy, Rice University, Houston, Texas 77005, USA*

[3]*Collaborative Innovation Center of Advanced Microstructures, Nanjing 210093, China*

[4]*Shanghai Research Center for Quantum Sciences, Shanghai 201315, China*

## Abstract

One main theme in current condensed matter physics is the search of quantum spin liquid (QSL), an exotic magnetic state with strongly-fluctuating and highly-entangled spins down to zero temperature without static order. However, there is no consensus on the existence of a QSL ground state in any real material so far. The disorders and competing exchange interactions may prevent the formation of an ideal QSL state on frustrated spin lattices. Here we report systematic heat transport measurements on a honeycomb-lattice compound $BaCo_2(AsO_4)_2$, which manifests magnetic order in zero field. In a narrow field range after the magnetic order is nearly suppressed by an in-plane field, in both perpendicular and parallel to the zigzag direction, a finite residual linear term of thermal conductivity is clearly observed, which is attributed to the mobile fractionalized spinon excitations. This provides smoking-gun evidence for a gapless QSL state in $BaCo_2(AsO_4)_2$. We discuss the underlying physics to form this exotic gapless QSL state in $Co^{2+}$ honeycomb lattice.

## Introduction

Quantum spin liquids (QSLs) are novel highly entangled quantum phases of matter which host fractionalized excitations[1-3]. In contrast to conventional ordered magnets, QSLs are characterized by the absence of long-range magnetic order and spontaneous symmetry breaking down to absolute zero temperature. Frustrations are thought to play a significant role in inducing quantum fluctuation and stabilizing QSL state[1-3]. Many geometrically frustrated systems have been considered as potential QSL candidates, such as prominent triangular-lattice compounds κ-(BEDT-TTF)$_2$Cu$_2$(CN)$_3$ (ref. [4-6]), EtMe$_3$Sb[Pd(dmit)$_2$]$_2$ (ref. [7-8]), YbMgGaO$_4$ (ref. [9-11]), NaYbSe$_2$ (ref. [12]), Kagome-lattice ZnCu$_3$(OH)$_6$Cl$_2$ (ref. [13]) and pyrochlore-lattice Tb$_2$Ti$_2$O$_7$, Pr$_2$Zr$_2$O$_7$, Yb$_2$Ti$_2$O$_7$, Ce$_2$Zr$_2$O$_7$, Ce$_2$Sn$_2$O$_7$ (ref. [14-17]). On the other hand, Kitaev proposed a two-dimensional honeycomb-lattice model in 2006, in which frustrations origin from the bond-dependent Ising-type interactions[18]. The exactly solvable Kitaev model hosts localized Z$_2$ gauge fluxes and itinerant Majorana fermions, the ground state of which could be three gapped QSL states (A phases) or one gapless QSL state (B phase), depending on the magnitude relationship among the three nearest-neighbor anisotropic interactions of each site[18]. Intriguingly, by introducing a magnetic field, the B phase will convert into a gapped non-Abelian phase with anyonic excitations obeying non-trivial braiding statistics, which could be utilized to realize intrinsically fault-tolerant topological quantum computations[19,20].

In the pursuit of Kitaev QSL in real materials, transition metal compounds with partially filled 4$d$ or 5$d$ shells were proposed as promising candidates due to the intricate interplay between electronic correlation and spin-orbit coupling (SOC)[21]. With strong SOC, the spatially anisotropic and bond-directional orbitals naturally lead to the Ising-type Kitaev interactions[22]. Along these lines, a flurry of studies have focused on Mott insulators of $d^5$ ions ($t_{2g}^5$ electronic configuration with pseudospin-1/2 Kramers doublet) with honeycomb lattice, including α-A$_2$IrO$_3$ (A= Li, Na, Cu), A$_3$LiIr$_2$O$_6$ (A=H, Ag, Cu) and α-RuCl$_3$ (ref. [23]). Nevertheless, except for H$_3$LiIr$_2$O$_6$, these materials all present long-range magnetic order at low temperature because of inevitable perturbations, such as conventional Heisenberg interactions $J$ resulting from direct $d$-$d$ hybridization, off-diagonal symmetric anisotropy $\Gamma$ originating from both $d$-$d$ and anion mediated $d$-$p$ electron transfer and additional $\Gamma'$ term due to trigonal distortion of ligand octahedra in real materials[24]. Applying external tuning parameter like magnetic field is an effective way to suppress the non-Kitaev terms and thus approaching the Kitaev QSL state[25,26]. Therefore, finding materials with dominant Kitaev interaction and negligible non-Kitaev terms is crucial in the search of Kitaev QSL.

Recent theoretical studies have proposed that 3$d^7$ Co$^{2+}$ ions with honeycomb lattice is another potential platform to host Kitaev QSL[27-30]. In an octahedral crystal field environment, Co$^{2+}$ ions form a $t_{2g}^5 e_g^2$ electronic configuration with a high spin state ($S$ = 3/2, $L$ = 1), thus resulting in pseudospin-

1/2 ground state Kramers doublet via SOC. Compared with weakly localized 4$d$ and 5$d$ systems, 3$d$ systems possess more compact $d$ orbitals, namely smaller long-range Heisenberg couplings. Furthermore, in despite of relatively weak SOC, the orbital moments remain active[29,30] and the additional $e_g$ orbitals-related exchange processes are ferromagnetic, which compensate the antiferromagnetic Heisenberg interactions from $t_{2g}$ orbitals and finally contribute to a dominant Kitaev coupling[27,28]. Experimentally, various cobaltates have been synthesized and investigated, such as $BaCo_2(PO_4)_2$ (ref. [31,32]), $BaCo_2(AsO_4)_2$ (BCAO) (ref. [33-35]), $Na_2Co_2TeO_6$ (ref. [36-38]) and $Na_3Co_2SbO_6$ (ref. [39,40]). Although these materials all undergo a magnetic transition at low temperature, interestingly, BCAO was proposed to host an intermediate QSL regime by recent thermodynamic[34] and THz spectroscopy experiments[35].

In this paper, with the motivation to identify QSL state in BCAO, we performed ultralow-temperature thermal conductivity measurements, a powerful technique to detect low-energy quasiparticles, on high-quality BCAO single crystals down to 80 mK with in-plane magnetic field. When the magnetic order is nearly suppressed by a weak field of $\mu_0 H \sim 0.5$ T, a finite residual linear term $\kappa_0/T$ shows up, which demonstrates the existence of mobile fractionalized spinon excitations. This spinon contribution of $\kappa_0/T$ persists in a narrow field range in both field directions (perpendicular and parallel to the zigzag direction), before the spins are more and more polarized. These results identify a gapless QSL state in BCAO, which emerges when part of the spins is still ordered, and vanishes when the spins are gradually polarized.

## Results

**Magnetic phase diagram.** In BCAO, edge-sharing $CoO_6$ octahedra form a two-dimensional (2D) layered honeycomb lattice which is stacked along the $c$ axis with an ABC periodicity. As sketched in Fig. 1a, $K_x$, $K_y$ and $K_z$ denote the bond-dependent Kitaev interactions between the nearest-neighborhood $Co^{2+}$ ions. The crystallographic $a$ axis is along the zigzag chain while $b$ axis is along the armchair direction, corresponding to the $[11\bar{2}]$ and $[\bar{1}10]$ directions in the (x,y,z) spin-axis coordinate, respectively. The Curie-Weiss temperature of BCAO is $\Theta_c = -167.7$ K and $\Theta_{ab} = 33.8$ K, respectively, indicating interplane antiferromagnetic and in-plane ferromagnetic exchange couplings[34], like $\alpha$-$RuCl_3$. Unlike $\alpha$-$RuCl_3$, no stacking faults or structural domains are detected in BCAO.

Figure 2a plots the magnetization from 2 to 20 K of BCAO in various magnetic fields parallel to the *b* axis. A clear peak due to antiferromagnetic (AFM) transition is observed at $T_{N2}$ = 5.4 K at zero field and it gradually shifts towards lower temperatures and eventually vanishes at about 0.53 T. Moreover, the magnetizations display a board hump $T_{N1}$ within the magnetic field range of 0.13 to 0.21 T. The upper and lower bounds indicate two successive metamagnetic transitions and realignments of in-plane spins. The first transition is from the double zigzag state to spin spiral state, the magnetic structures of which are illustrated in Fig. 1b and 1c (ref. [33]). Thereinto, the spin spiral state presents weak ferromagnetic correlation, evidenced by the hysteresis loop in Fig. 2b. With field increasing, the system will then undergo a second transition into the collinear AFM state, corresponding to the nearly 1/3 plateau of the saturated magnetization, as shown in Fig. 2b. As a comparison, the magnetization in fields along *a* axis was also measured (See Supplementary Note 2), which illustrates weak in-plane anisotropy except that the polarization field is slightly higher in the *a* axis case.

The magnetic phase diagram summarized in Fig. 2c is roughly consistent with the previous report[34]. The phase boundaries below 0.7 K are determined from our thermal conductivity measurements, which will be discussed later. Most importantly, in this work we will demonstrate that there exist a gapless QSL regime after the AFM order is nearly suppressed, characterized by the finite residual linear term of thermal conductivity.

**Field dependence of thermal conductivity.** The in-plane ultralow-temperature thermal conductivity of BCAO Sample A in zero and finite magnetic fields up to 8 T is displayed in Fig. 3a-3e. The heat current is along *a* axis while the field is along *b* axis, as shown in the inset of Fig. 3a. Figure 3g and 3h plot the field dependence of $\kappa/T$ at several temperatures, and these isotherms exhibit almost the same behavior. As the magnetic field increases, $\kappa/T$ first drops slightly until $\mu_0 H \sim$ 0.05 T, then increases sharply for $\mu_0 H <$ 0.2 T, followed by a second decrease until $\mu_0 H \sim$ 0.53 T, then rises monotonously and finally saturates for $\mu_0 H >$ 5 T. The critical fields are consistent with above magnetization results and the whole evolution trend is consistent with previous thermal conductivity results in the temperature range of 4 to 4.8 K (ref. [34]).

The first two extreme points in Fig. 3h at $H_{c1}$ (0.05 T) and $H_{c2}$ (0.2 T) clearly indicate two successive field-induced alternations of magnetic structure in BCAO, which define the boundaries of the spin spiral state. The least value occurs at $H_{c3}$ (0.53 T), implying strong quantum fluctuations when the AFM order is suppressed. In the spin spiral state, owing to the weak ferromagnetic correlation, the magnons are increasingly gapped out[41], which weakens the spin-phonon scattering and leads to the enhancement of $\kappa$. In the polarized state, the suppression of spin-phonon scattering due to the alignment of spins with increasing magnetic field results in the monotonical increase of $\kappa$. Since the thermal conductivity $\kappa$ at 5 T and 8 T nearly overlap with each other, it means that $\kappa$ above 5 T is entirely contributed by phonons without scattering from magnetic system in the fully polarized state. Below we focus on the residual linear term of thermal conductivity in different states.

**Absence of residual linear term $\kappa_0/T$ in magnetically ordered states and spin polarized state.** In Fig. 3f, we fit the thermal conductivity data below 0.25 K for $\mu_0H$ = 0, 0.1, 0.2, 5T and 8 T to $\kappa/T = a + bT^{\alpha-1}$, where the two terms $aT$ and $bT^{\alpha}$ represent contributions from itinerant fermionic excitations and phonons, respectively[42,43]. For phonons, the power $\alpha$ is typically between 2 and 3, owing to the specular reflections at the sample surfaces[42,43]. Note that if gapless AFM spin waves (magnons) exist, their contributions behave as $T^3$ and will not blur our analysis of residual linear term[44]. The fitting yields $\kappa_0/T \equiv a$ = 0.004 ± 0.017 mW K$^{-2}$ cm$^{-1}$ for 0 T, 0.004 ± 0.023 mW K$^{-2}$ cm$^{-1}$ for 0.1 T, -0.005 ± 0.025 mW K$^{-2}$ cm$^{-1}$ for 0.2 T, 0.002 ± 0.020 mW K$^{-2}$ cm$^{-1}$ for 5 T, and $\kappa_0/T$ = -0.003 ± 0.027 mW K$^{-2}$ cm$^{-1}$ for 8 T, respectively. Considering our experimental error bar ±0.005 mW K$^{-2}$ cm$^{-1}$, the $\kappa_0/T$ of BCAO at these magnetic fields is virtually zero. The absence of $\kappa_0/T$ is reasonable in these magnetically ordered states and spin polarized state, because magnetic excitations with a gap larger than 1 meV were observed in the INS experiments[41]. In fact, what we care most is whether $\kappa_0/T$ shows up in the intermediate regime when the magnetic order is suppressed.

**Mobile gapless spinon excitations in the intermediate regime.** In Fig. 4a, we fit the data from 0.4 T to 0.65 T below 0.25 K to $\kappa/T = a + bT^{\alpha-1}$. Although the fittings of $\kappa/T$ extrapolating to zero at the zero-temperature limit exhibit pure phonon contributions, i.e. with zero $\kappa_0/T$, for $\mu_0H$ = 0.4, 0.45, 0.6, and 0.65 T, strikingly, finite $\kappa_0/T$ is obtained between 0.45 and 0.6 T. To see more details, we plot the thermal conductivity data for $\mu_0H$ = 0.5, 0.53, and 0.55 T in Fig. 4b. The fittings give $\kappa_0/T$ = 0.050 ± 0.007 mW K$^{-2}$ cm$^{-1}$, 0.058 ± 0.004 mW K$^{-2}$ cm$^{-1}$, and 0.055 ± 0.004 mW K$^{-2}$ cm$^{-1}$, respectively.

To check the reproducibility of this result, the thermal conductivity of three more samples was also measured and the results at 0.55 T are plotted in Fig. 4c. The fittings yield $\kappa_0/T$ = 0.044 ± 0.002 mW K$^{-2}$ cm$^{-1}$ for Sample B, 0.033 ± 0.005 mW K$^{-2}$ cm$^{-1}$ for Sample C, and $\kappa_0/T$ = 0.033 ± 0.003 mW K$^{-2}$ cm$^{-1}$ for sample D, which shows that our results are highly reproducible.

These nonzero $\kappa_0/T$ immediately imply that there exists a quantum spin liquid state accompanied by mobile gapless fermionic excitations in BCAO. Indeed, the $T$-linear behavior in $\kappa_{xx}$ at the zero-temperature limit has been theoretically predicted before by itinerant Majorana fermions on the honeycomb lattice[45]. According to the method in ref. [7], assuming the fermionic excitations are in analogy with electrons near the Fermi surface in metals, and with linear dispersion relation, the mean free path ($l_s$) can be estimated by calculating $\frac{\kappa_0}{T} = \frac{\pi k_B^2}{9h} \frac{l_s}{a'd} = \frac{\pi}{9}(\frac{k_B}{h})^2 \frac{J}{d}\tau_s$. Here $a' = \frac{a}{\sqrt{3}} = 2.89$ Å and $d \approx \frac{c}{3} = 7.83$ Å represent the nearest-neighbor in-plane and out-of-plane Co-Co distance in BCAO, respectively[34]. From the observed $\kappa_0/T$ = 0.055 mW K$^{-2}$ cm$^{-1}$ for $\mu_0H$ = 0.55 T, the obtained $l_s$ is 123.8 Å, which is about 43 times the in-plane interspin distance without being scattered, indicating the mobile nature of the fermionic excitations.

For comparison, the thermal conductivity of Sample B with field along $a$ axis is also measured, as plotted in Fig. 4e. The fittings below 0.25 K give $\kappa_0/T$ = 0.025 ± 0.002 mW K$^{-2}$ cm$^{-1}$ for 0.5 T, 0.034 ± 0.005 mW K$^{-2}$ cm$^{-1}$ for 0.55 T, 0.028 ± 0.004 mW K$^{-2}$ cm$^{-1}$ for 0.6 T, respectively. Therefore, a finite $\kappa_0/T$ is observed in both in-plane field directions. Interestingly, the fitting gives $\kappa_0/T$ = -0.002 ± 0.017 mW K$^{-2}$ cm$^{-1}$ with α = 1.94 ± 0.08 for 0.65 T. The power is abnormally lower than 2, which is similar to YbMgGaO$_4$[11]. The case for 0.75 T is the same as 0.65 T, whose fitting gives $\kappa_0/T$ = -0.139 ± 0.026 mW K$^{-2}$ cm$^{-1}$ with α = 1.68 ± 0.04. A negative $\kappa_0/T$ has no physical meaning, and this sublinear behavior can be attributed to strong scattering by magnetic fluctuations[11].

The field dependence of $\kappa_0/T$ is plotted in Fig. 4d and 4f for $H \parallel b$ (Sample A) and $H \parallel a$ (Sample B), respectively. The maximum $\kappa_0/T$ of Sample B in $\mu_0H$ = 0.55 T $\parallel a$ is relatively smaller, with calculated $l_s \sim 76.5$ Å, about 26 times as long as the in-plane interspin distance. Nevertheless, the field range of gapless QSL state is relatively wider for $H \parallel a$, which may link to the slightly higher polarization field along $a$ axis. Note that the gapless QSL state has some overlap with both collinear AFM state and partially polarized state, as seen in the phase diagram of Fig. 2d. A simple understanding is that the gapless QSL state can coexist with both partially ordered state and partially polarized state, and the fractionalized spinons of the QSL state are still mobile.

## Discussion

In the search of quantum spin liquids, detecting the emergent fractionalized spinon excitations is crucial and ultralow-temperature thermal conductivity measurement is one of the most low-energy experimental techniques. Previously triangular-lattice organic compound EtMe$_3$Sb[Pd(dmit)$_2$]$_2$ (ref. [7]) has been reported to exhibit a huge $\kappa_0/T$ of 2 mW K$^{-2}$ cm$^{-1}$, however, it cannot be reproduced by other groups[46,47]. To our knowledge, there is still no reproducible report of the existence of finite $\kappa_0/T$ in any QSL candidate so far, including honeycomb-lattice Kitaev materials such as α-RuCl$_3$ (ref. [48]). Therefore, it is striking that a finite $\kappa_0/T$ is observed in a narrow field range in BCAO. This is highly reproducible, in several samples and two field directions. Below we discuss the underlying physics to form this exotic gapless QSL state.

First, we should emphasize that our results are in sharp contrast to the gapless Kitaev spin liquid with a static Z$_2$ guage field (B phase). Within the framework of pure Kitaev model, the itinerant Majorana fermions are gapless with two Dirac nodes[18]. When applying a magnetic field, B phase will acquire a gap $\Delta_M \sim \frac{h_x h_y h_z}{K^2}$ (here, $h_i$ represents the Cartesian component of the applied field) in the bulk and thus forming a topologically protected gapless chiral Majorana edge mode[18]. Therefore, it is obvious that the bulk excitations remain gapless if at least one component disappears. Due to the particular geometry and oxygen-mediated hopping of edge-shared CoO$_6$ octahedra in BCAO, which are similar to iridates[22], the above condition can be realized when the magnetic field is applied along *b* axis, as depicted in Fig. 1a. In other words, the Majorana gap has six-fold symmetry with respect to the in-plane magnetic field. This in-plane field-angle dependence of physical quantities has been observed in α-RuCl$_3$ by magnetization[49], thermal hall conductivity[50], and specific heat[51] measurements. However, our results on ultralow-temperature thermal conductivity reveal the existence of gapless fermionic excitations when the field is applied in both *a* and *b* axes for BCAO, which is apparently beyond the hypothesis.

Another possible scenario is that the intermediate regime is a U(1) gapless quantum spin liquid with neutral spinon Fermi surfaces[52-54]. In this framework, the bulk gapless fermionic excitations will immediately open a gap when a magnetic field is applied, and then the system will undergo a quantum transition from the Z$_2$ gapped Kitaev QSL with non-Abelian Ising topological order to the intermediate U(1) gapless QSL as the field increases, and finally enter into a gapped trivial polarized state[52]. Especially, the gapless QSL state is stable even in the presence of additional small Heisenberg and off-diagonal gamma interactions[52]. This picture can be depicted by the analogy of fermiology of spinons. The gapped non-Abelian phase corresponds to a $p_x+ip_y$ chiral topological superconductor of fermionic spinons. Hence, applying a magnetic field will destroy the superconducting order and form a spinon metal with a gapless Fermi surface coupled to a massless U(1) gauge field[53]. Theoretically, considering the inevitable disorder of real samples, the U(1) gapless QSL system will exhibit a $T$-linear behavior in the in-plane thermal conductivity at the ultralow-temperature limit[55]. The prediction is consistent with our experimental observations. However, this scenario is based on dominate AFM Kitaev interaction, while the gapless QSL state is absent in the ferromagnetic case[52-54], which cannot reconcile with BCAO. Furthermore, the intermediate state has also been proposed as a gapped chiral spin liquid[56] or does not exist when magnetic field is applied along $b$ axis in 2D Kitaev honeycomb lattice[57,58]. These disagreements with our observation of robust $\kappa_0/T$ imply that more experimental and theoretical efforts are needed to determine the physical origin of the exotic gapless QSL state in BCAO. Interestingly, we notice that recent study demonstrated a novel mechanism to induce Majorana Fermi surfaces with U(1) degrees of freedom, independent of the sign of Kitaev interaction[59].

Finally, we cannot exclude the scenario that BCAO is described by XXZ-$J_1$-$J_3$ model rather than extended Kitaev model ($JK\Gamma\Gamma'$). Indeed, whether there exists large Kitaev coupling in Co-based layered honeycomb system is still under debate. Recent theoretical studies[60-63] and INS experiments[41] have favored that third nearest neighbor Heisenberg interaction $J_3$ plays a significant role in $3d^7$ cobaltates, leading to dominant FM Heisenberg hopping with negligible Kitaev coupling in BCAO, similar to BCPO (ref. [32]). The XXZ-$J_1$-$J_3$ model might also account for the small anisotropy in magnetizations observed in BCAO under two in-plane field orientations. Within this framework, the competition between $J_1$ and $J_3$ is a feasible mechanism to induce geometric frustration, thus approaching the gapless QSL state[41,64].

In summary, we have measured the ultralow-temperature thermal conductivity of honeycomb-lattice $BaCo_2(AsO_4)_2$ single crystals down to 80 mK with in-plane magnetic field up to 8 T. At finite temperatures, the field dependence of the thermal conductivity exhibits a series of extreme points, corresponding to successive metamagnetic transitions. At the zero-temperature limit, finite residual linear terms are clearly observed when the magnetic order is nearly suppressed, which strongly support a field-induced gapless QSL state with mobile spinon excitations. Moreover, the gapless QSL state is robust in magnetic fields along both $a$ and $b$ axes. These results exclude the possibility of gapless Kitaev QSL and put strong constraints on the theoretical description of $BaCo_2(AsO_4)_2$.

**Methods**

**Sample preparation and characterization.**

We prepare the high-quality single crystals of $BaCo_2(AsO_4)_2$ using the flux-method as described in ref. [34]. The x-ray diffraction (XRD) measurement was performed on a typical $BaCo_2(AsO_4)_2$ sample by using an x-ray diffractometer (D8 Advance, Bruker), and determined the largest surface to be the (001) plane (see Supplementary Material). The crystallographic axes were identified via Laue diffraction.

**DC magnetization and specific heat measurements**. The DC magnetization measurement was performed down to 2 K using a magnetic property measurement system (MPMS, Quantum Design). The specific heat data were measured on a 1.4 mg single crystal by the relation method down to 0.3 K using a physical property measurement system (PPMS, Quantum Design) equipped with a $^3$He cryostat.

**Thermal transport measurements.** The $BaCo_2(AsO_4)_2$ single crystals for thermal conductivity measurements were cut into rectangular shapes in the *ab* plane. The dimensions are $5.33 \times 0.95 \times 0.27$ mm$^3$, $5.06 \times 0.66 \times 0.26$ mm$^3$, $2.50 \times 0.75 \times 0.28$ mm$^3$, $3.89 \times 0.63 \times 0.42$ mm$^3$ for Sample A, B, C, and D, respectively. The thermal conductivity was measured in a dilution refrigerator, using a standard four-wire steady-state method with two $RuO_2$ chip thermometers, calibrated *in situ* against a reference $RuO_2$ thermometer. The heat current was along *a* axis in all measurements, while the magnetic fields were applied along either *a* or *b* axis.

## Data availability

The data that support the findings of this study are available from the corresponding authors upon reasonable request.

## Acknowledgements

We thank Tao Wu, Yi Zhou, and Hong Yao for helpful discussions. This research was funded by the National Natural Science Foundations of China (Grant No. 12034004 and No. 12174064), and the Shanghai Municipal Science and Technology Major Project (Grant No. 2019SHZDZX01). The single crystal growth and characterization efforts at Rice are supported by U.S. DOE BES DE-SC0012311 and the Robert A. Welch Foundation under Grant No. C-1839, respectively (P.D.).

## Author Contributions

S.Y.L. conceived the idea and designed the experiments. C.P.T and D.Z.D. performed the DC magnetization and bulk thermal transport measurements with help from X.Z., C.C.Z., X.B.J., B.G. , T.C. and P.C.D. synthesized the single crystal samples. C.P.T. and S.Y.L. wrote the manuscript with comments from all authors. C.P.T. and D.Z.D. contributed equally to this work.

## Competing interests

The authors declare no competing interests.

## Additional Information

**Supplementary information** is available for this paper at URL inserted when published.

**Correspondence** and requests for materials should be addressed to S.Y.L. (shiyan_li@fudan.edu.cn).

Figure 1

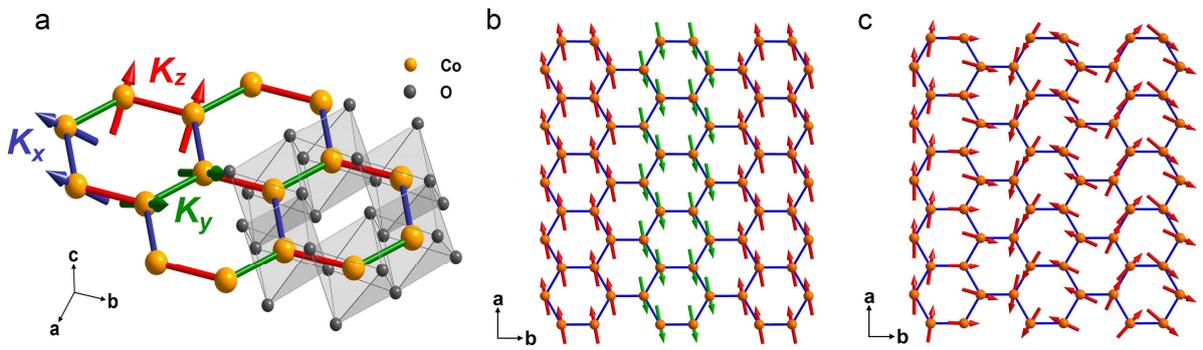

Figure 2

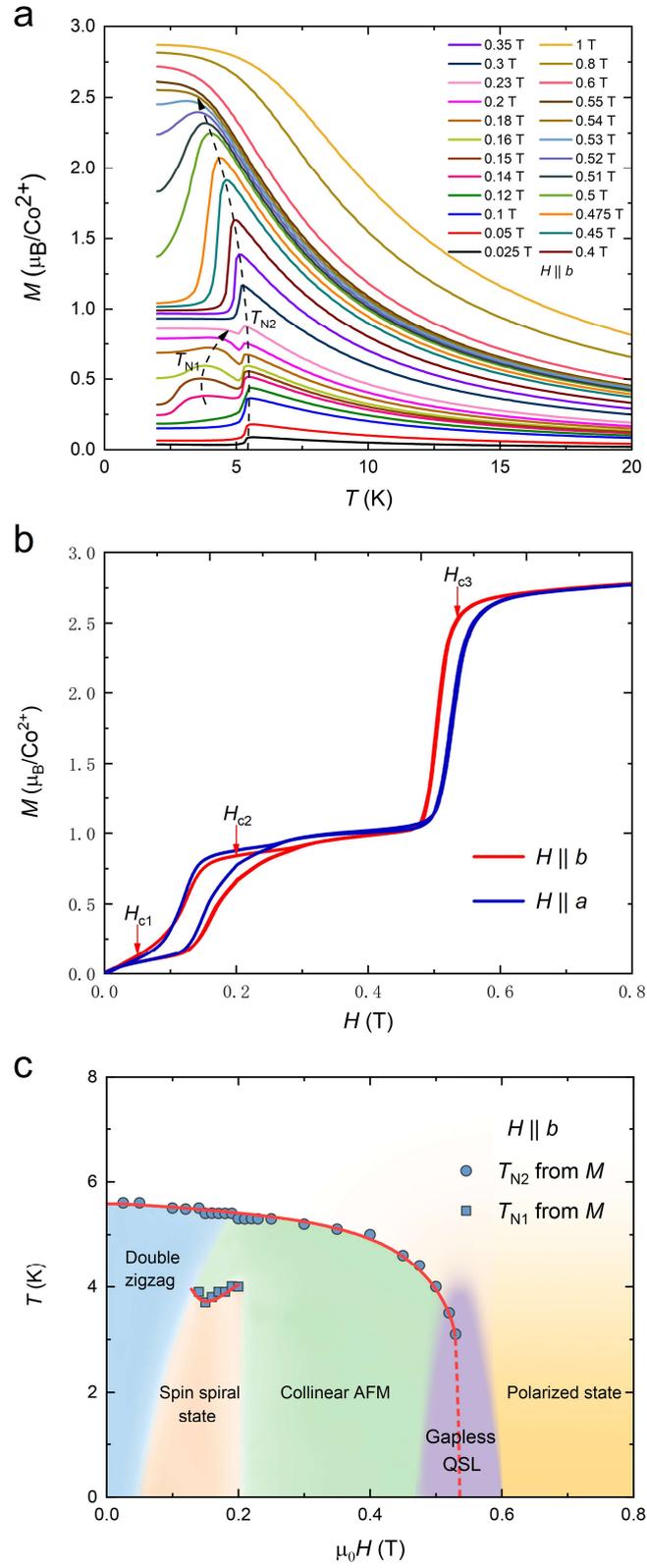

Figure 3

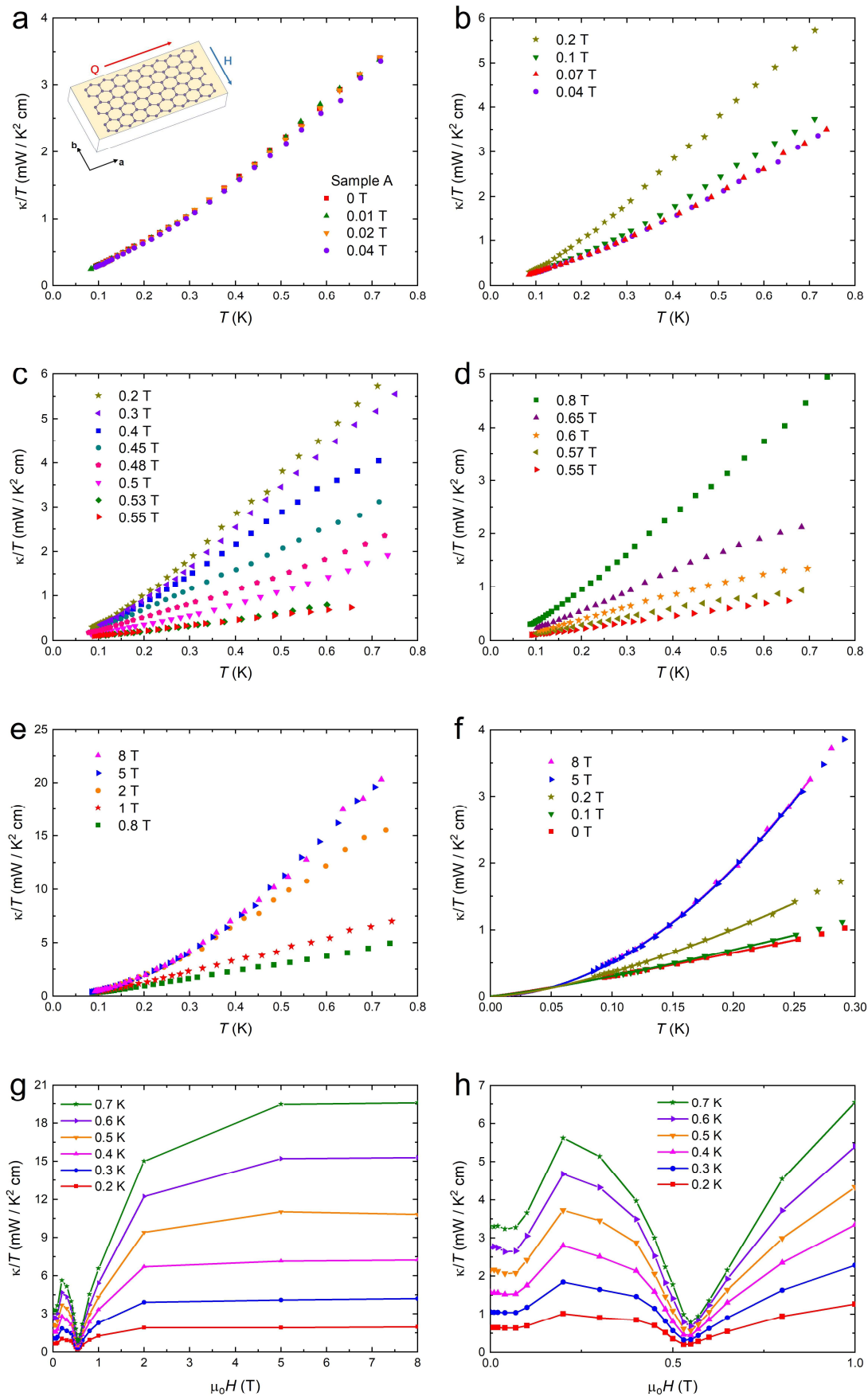

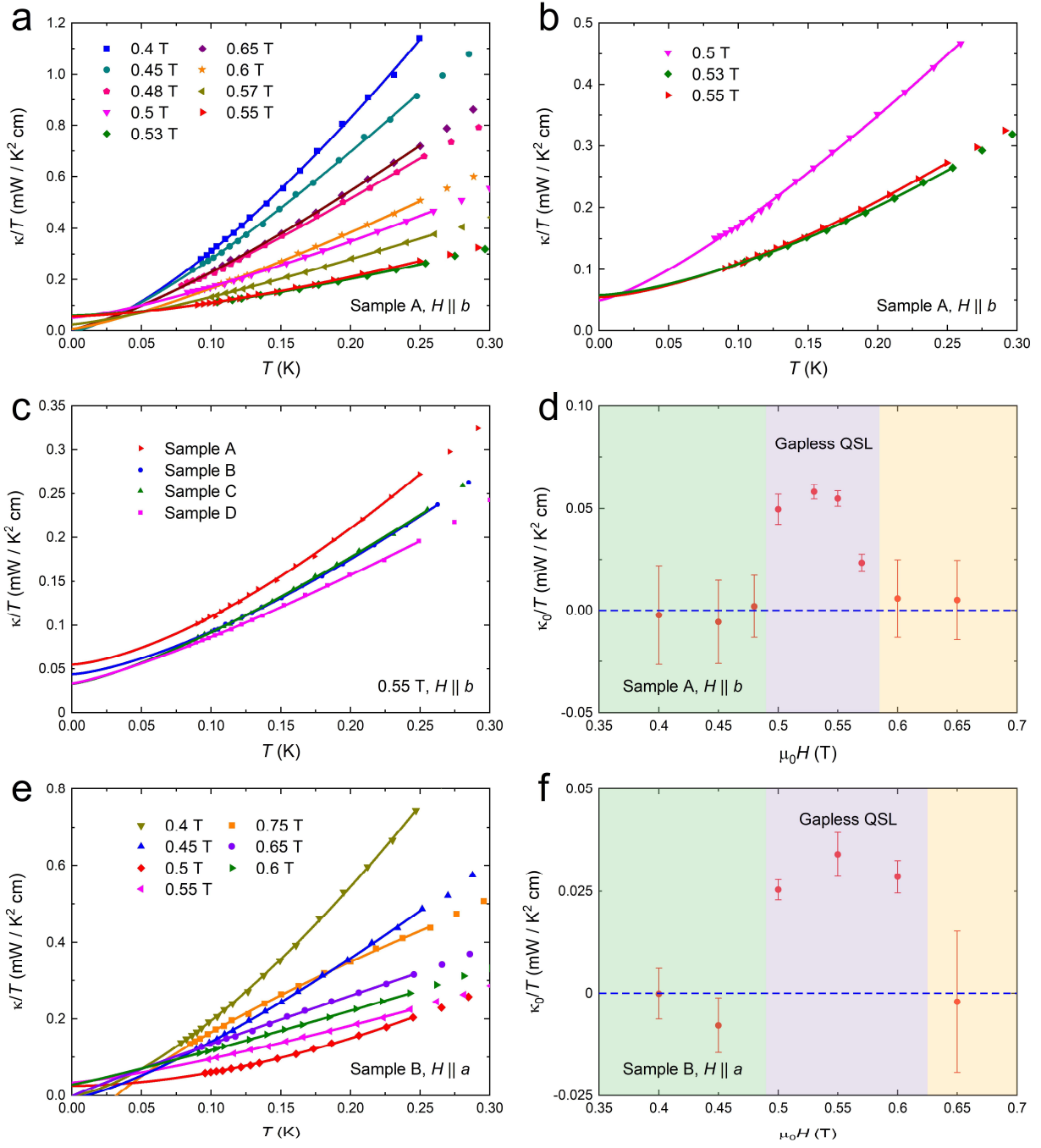

**Figure captions**

**Figure 1 | Crystal structure and magnetic structure of $BaCo_2(AsO_4)_2$.** (a) Crystal structure of $BaCo_2(AsO_4)_2$ in *ab* plane and definitions of crystallographic axes *a*, *b*, *c* as well as the spin axes *x*, *y*, *z*. Spin *x*, *y*, *z* axes, perpendicular to Co-Co bond, are determined by bond-dependent Kitaev interaction $K_x$, $K_y$, $K_z$. Yellow and grey circles represent $Co^{2+}$ and $O^{2-}$, respectively. (b) Double zigzag spin-chains form a ↑↑↓↓ pattern with small out-of-plane canting angle. (c) In-plane helical structure of spin spiral state, showing the stacking of weak coupled quasi-ferromagnetic chains[33].

**Figure 2 | Weak in-plane field manipulation of the magnetic structure and the phase diagram for $H \parallel b$.** (a) Temperature-dependent magnetic susceptibilities of $BaCo_2(AsO_4)_2$. The dashed arrows indicate two phase transitions at $T_{N1}$ and $T_{N2}$ for $H \parallel b$. (b) Field-dependent magnetization for fields along both *a* and *b* axes at 2 K, showing small in-plane anisotropy. The critical fields extracted from magnetization (Fig. 2b) and thermal conductivity (Fig. 3g and 3h) are labeled as $H_{c1}$, $H_{c2}$ and $H_{c3}$ with red arrows. (c) The phase diagram for $H \parallel b$, exhibiting the evolution between several magnetic ordered states, gapless QSL state, and polarized state. The Néel temperature $T_{N1}$ and $T_{N2}$ are determined from magnetizations in Fig. 2b.

**Figure 3 | Heat transport results of $BaCo_2(AsO_4)_2$ for fields along *b* axis.** The thermal conductivity $\kappa$ of Sample A at various applied fields of 0 - 0.04 T (a), 0.04 - 0.2 T (b), 0.2 - 0.55 T (c), 0.55 - 0.8 T (d), and 0.8 - 8 T (e). The inset in (a) shows the configuration of $Q \parallel a$ and $H \parallel b$, here $Q$ represents the heat current. (f) Thermal conductivity for several fields in magnetic ordered states and polarized state, plotted as $\kappa/T$ vs. $T$. The solid lines represent fittings by $\kappa/T = a + bT^{\alpha-1}$ below 0.25 K. The values of $\kappa_0/T$ are negligible for these fields. (g) and (h) Field dependence of thermal conductivity plotted within two field ranges. Note that the extreme points, corresponding to $H_{c1}$, $H_{c2}$, and $H_{c3}$, are consistent with magnetization results, which indicate several magnetic phase transitions.

**Figure 4 | Heat transport results at various fields around the gapless QSL region.** (a) Thermal conductivity in magnetic fields from 0.4 to 0.55 T for Sample A. The solid lines represent fittings below 0.25 K, the same as in Fig. 3. The data with finite residual linear term are replotted in (b). (c)

Thermal conductivity of four different samples at 0.55 T. All the data yield finite residual linear terms, manifesting the highly reproducibility of our results. (d) Field dependence of the $\kappa_0/T$. The finite $\kappa_0/T$ exists within the field range from about 0.5 to 0.6 T, which outlines the boundary of gapless QSL state. (e) Thermal conductivity of Sample B for fields along $a$ axis. Finite $\kappa_0/T$ is observed at several fields illustrated in (f), identifying the robust gapless QSL state for magnetic field along both $a$ and $b$ axes.

# Supplementary Information for "Evidence for gapless quantum spin liquid in a honeycomb lattice"

Chengpeng Tu[1], Dongzhe Dai[1], Xu Zhang[1], Chengcheng Zhao[1], Xiaobo Jin[1], Bin Gao[2], Tong Chen[2], Pengcheng Dai[2], and Shiyan Li[1,3,4*]

[1]State Key Laboratory of Surface Physics, and Department of Physics, Fudan University, Shanghai 200438, China

[2]Department of Physics and Astronomy, Rice University, Houston, Texas 77005, USA

[3]Collaborative Innovation Center of Advanced Microstructures, Nanjing 210093, China

[4]Shanghai Research Center for Quantum Sciences, Shanghai 201315, China

# Supplementary Note 1: X-ray diffraction measurement

The typical x-ray diffraction (XRD) data is plotted in Supplementary Fig. 1, determining the largest surface of the samples to be the (001) plane. The heat current and magnetic fields are applied in the (111) plane. The full width at half maximum (FWHM) is only 0.096°, indicating the high quality of our samples.

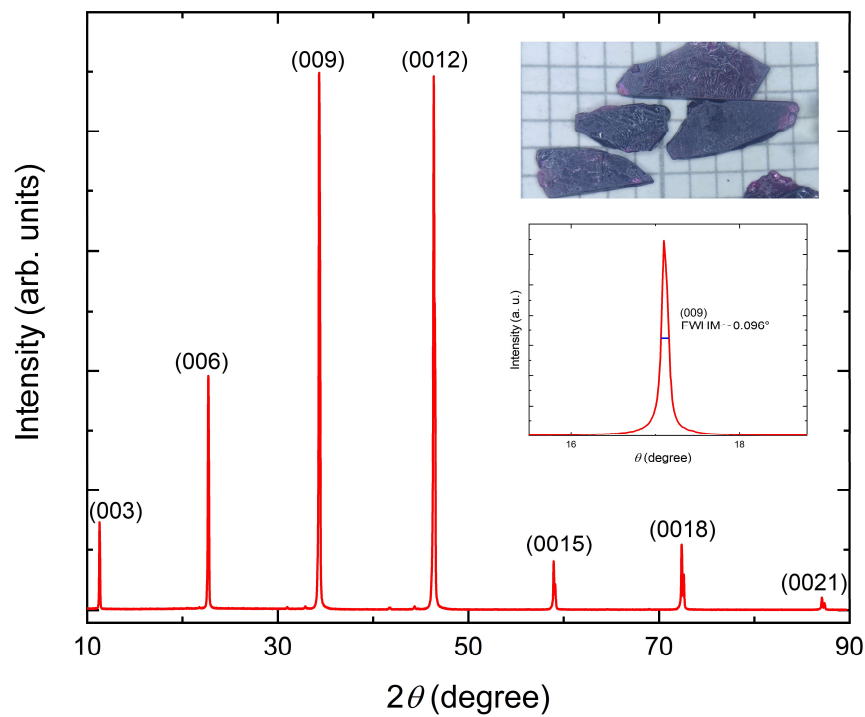

**Supplementary Figure 1 | Typical x-ray diffraction pattern of the $BaCo_2(AsO_4)_2$ single crystal.** The largest plane is determined to be the (001) plane.

## Supplementary Note 2: Magnetization for $H \parallel a$

Supplementary Fig. 2 depicts the magnetization from 2 to 20 K of BaCo$_2$(AsO$_4$)$_2$ single crystal in various magnetic fields parallel to the $a$ axis. The whole behavior is similar to the case where fields are applied along the $b$ axis, indicating weak in-plane anisotropy. The magnetic order vanishes at about 0.54 T, slightly larger than the critical field $H_{c3}$ (0.53 T) for $H \parallel b$, which can also be evidenced by magnetizations in Fig. 2b. Besides, the range of the spin spiral state is larger, from 0.11 to 0.26 T.

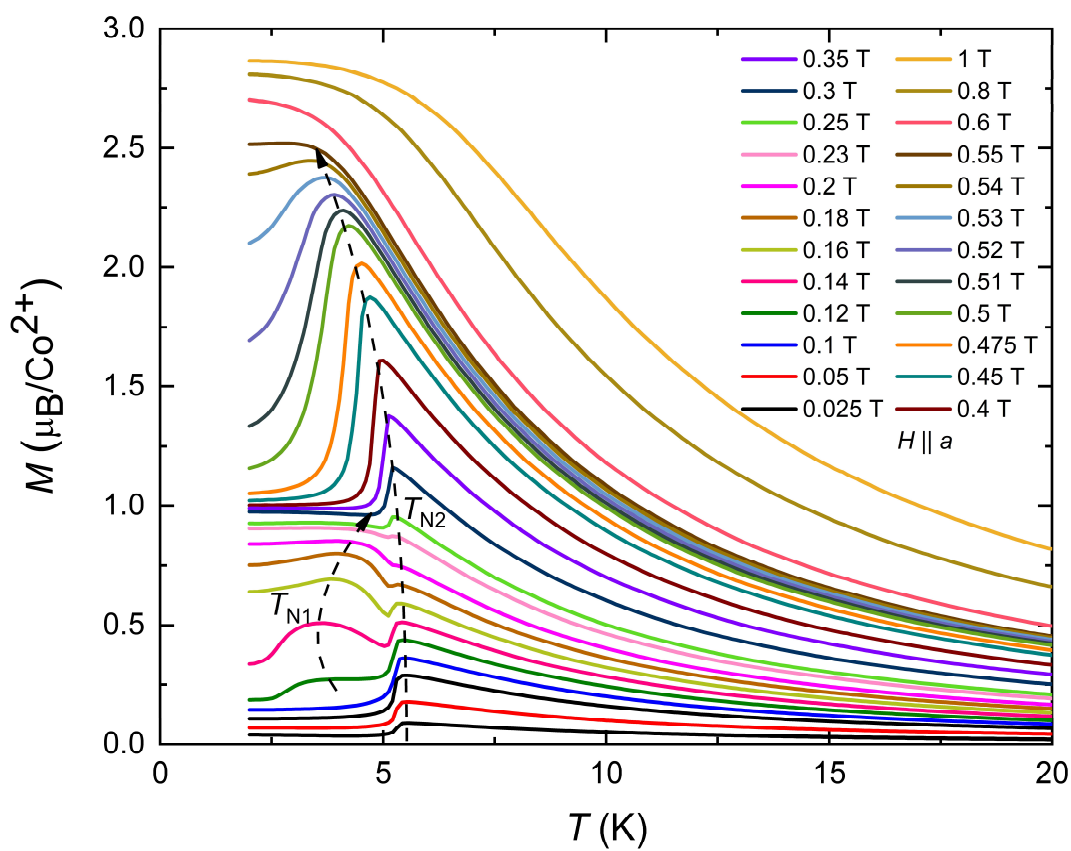

**Supplementary Figure 2 | Low-temperature magnetization of the BaCo$_2$(AsO$_4$)$_2$ single crystal at various magnetic fields along $a$ axis.**

## Supplementary Note 3: Low-temperature specific heat measurements for $H \parallel a$

The temperature dependence of specific heat for $H \parallel a$ is plotted in Supplementary Fig. 3(a). At zero field, a λ-peak due to antiferromagnetic ordering is observed at $T_{N2} = 5.15$ K, which is then suppressed by a small in-plane field above 0.6 T. It is roughly consistent with the previous specific heat measurements[S1,S2] and our magnetization results. As temperature decreases, the values of $C_p/T$ exhibit a sudden upturn below about 1 K, which can be attributed to a high-temperature tail of a nuclear Schottky peak. To see this more clearly, we replot $C_p/T$ below 4 K in Supplementary Fig. 3(b).

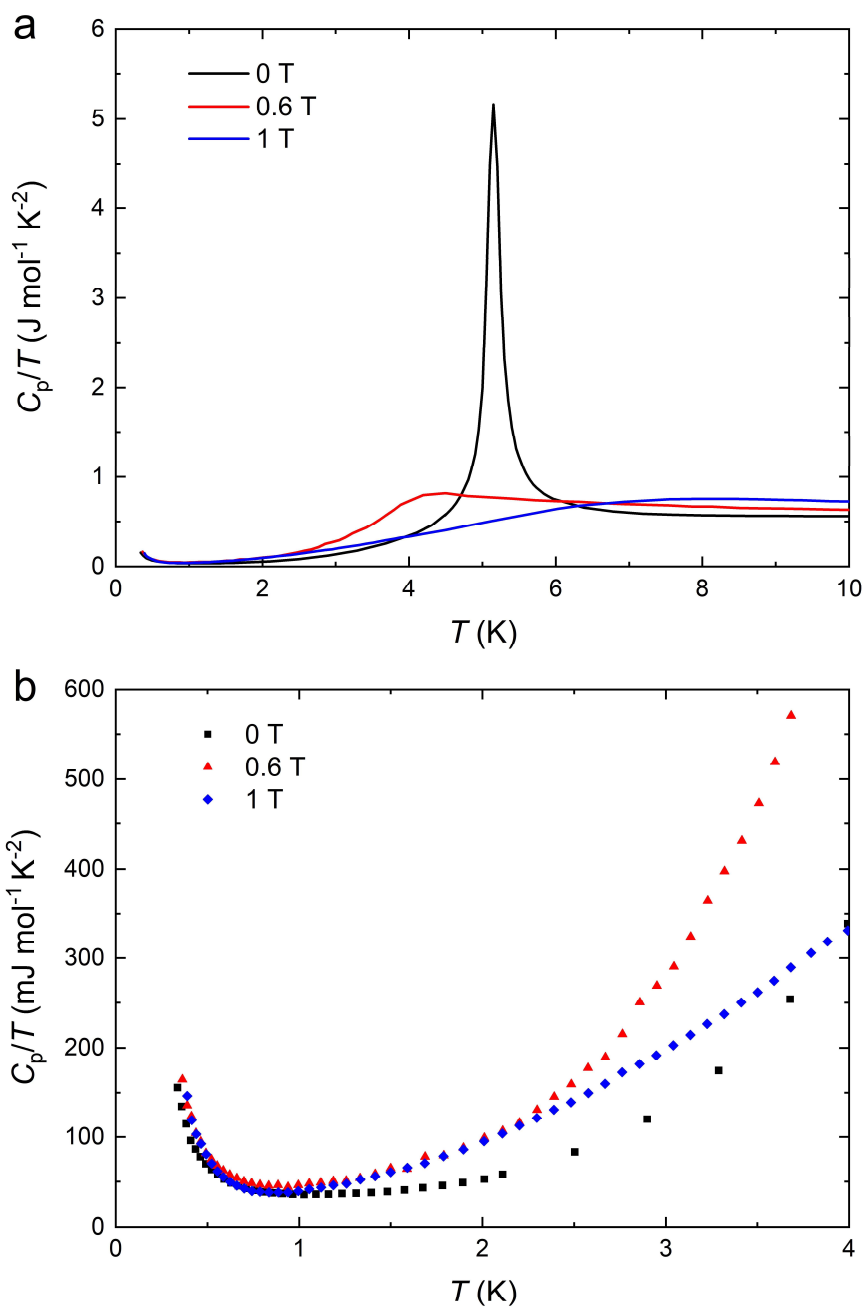

**Supplementary Figure 3 | The low-temperature specific heat of the BaCo$_2$(AsO$_4$)$_2$ single crystal for *H* || *a*.** (a) The temperature dependence of $C_p/T$ below 10 K at several fields. At zero field, a clear phase transition occurs at $T_{N2}$ = 5.15 K. (b) An abrupt upturn is observed below about 1K, which can be attributed to the nuclear Schottky contribution.

## Supplementary References